\documentclass[fleqn, twoside]{article}
\usepackage{graphics}
\begin{document}
\title{Black holes: A physical route to the Kerr metric}
\author{R.\ Meinel\\ University of Jena,
Institute of Theoretical Physics,\\
Max-Wien-Platz  1, 07743 Jena, Germany}
\date{meinel@tpi.uni-jena.de}
\maketitle
\begin{abstract}
As a consequence of Birkhoff's theorem, the exterior gravitational
field of a spherically symmetric star or black hole is always given
by the Schwarzschild metric. In contrast, the exterior gravitational
field of a rotating (axisymmetric) star differs, in general,
from the Kerr metric, which describes a stationary, rotating black hole.

In this paper I discuss the possibility of a quasi--stationary transition
from rotating equilibrium configurations of normal matter to rotating
black holes.
\end{abstract}
\section{Introduction: The Kerr black hole}
\label{intro}
The Kerr metric \cite{kerr}, in Boyer--Lindquist coordinates \cite{bl}, is
given by
\begin{equation}
ds^2=\Sigma\,(\frac{dr^2}{\Delta}+d\vartheta^2)+W^2 e^{-2\nu}(d\varphi-\omega\,
dt)^2-e^{2\nu}dt^2
\label{k}
\end{equation}
with
\begin{equation}
\Sigma=r^2+a^2\cos^2\vartheta, \quad \Delta=r^2-2Mr+a^2, \quad
W^2=\Delta\sin^2\vartheta,
\end{equation}
\begin{equation}
e^{2\nu}=\frac{\Delta\Sigma}{(r^2+a^2)^2-\Delta a^2\sin^2\vartheta} \quad
\mbox{and}\quad \omega=\frac{2Mra}{\Delta\Sigma}\,e^{2\nu}.
\end{equation}
It depends on two parameters, the total mass $M$ and the angular momentum
$J=Ma$ (we assume $J\ge 0$ without loss of generality, and we use units
where the velocity of light $c$ as well as Newton's gravitational constant
$G$ are equal to 1). The metric is stationary (independent of $t$) and
axisymmetric (independent of $\varphi$). The horizon of the black hole is
given by
\begin{equation}
r=r_+\equiv M+\sqrt{M^2-a^2},
\end{equation}
the larger root of the quadratic equation $\Delta=0$. Note that the Kerr
metric describes a black hole only if
\begin{equation}
a\le M \quad (J\le M^2)
\label{jmk}
\end{equation}
is satisfied. The boundary of the `ergosphere' is characterized by
\begin{equation}
r=r_0(\vartheta)\equiv M+\sqrt{M^2-a^2\cos^2\vartheta}.
\label{ergo}
\end{equation}
Within the ergosphere ($r_+<r<r_0$) any observer must rotate in the same
direction as the black hole ($d\varphi/dt>0$).

It is interesting to discuss circular orbits of test particles in the
`equatorial plane' $\vartheta=\pi/2$. Their angular velocity is given by
\begin{equation}
\Omega=\pm \frac{\sqrt{M}}{r^{3/2}\pm a\sqrt{M}}\quad ,
\end{equation}
where the upper sign characterizes direct orbits
(corotating with the black hole) and the lower sign holds for retrograde
(counterrotating) orbits. The circular orbits exist only for $r>r_{\rm ph}$,
with the `photon orbit'
\begin{equation}
r_{\rm ph}=2M\left\{1+\cos\left[\frac{2}{3}\arccos
\left(\mp\frac{a}{M}\right)\right]\right\}.
\end{equation}
The orbits are bound for $r>r_{\rm mb}$, with the `marginally bound orbit'
\begin{equation}
r_{\rm mb}=2M\mp a+2M^{1/2}(M\mp a)^{1/2}.
\end{equation}
(A particle in an unbound orbit will, under the influence of an
infinitesimal outward perturbation, escape to infinity.)
The orbits are stable for $r>r_{\rm ms}$,
with the `marginally stable orbit'
\begin{equation}
r_{\rm ms}=M\left\{3+Z_2\mp \left[(3-Z_1)(3+Z_1+2Z_2)\right]^{1/2}\right\}
\end{equation}
\begin{equation}
\mbox{where} \quad
Z_1=1+\left(1-\frac{a^2}{M^2}\right)^{1/3}\left[\left(1+\frac{a}{M}\right)^{1/3}
+\left(1-\frac{a}{M}\right)^{1/3}\right]
\end{equation}
\begin{equation}
\mbox{and} \quad Z_2=\left(3\frac{a^2}{M^2}+Z_1^2\right)^{1/2}.
\end{equation}
These results on circular orbits of test particles
were derived by Bardeen {\it et al.} \cite{bpt}, see also \cite{st}.
\section{Limiting cases}
The two limiting cases of the Kerr black hole are $a=0$ ($J=0$), the
nonrotating (Schwarz\-schild) black hole, and $a=M$ ($J=M^2$), the
maximally rotating (extreme Kerr) black hole. For $a=0$ the horizon is given
by $r_+=2M$ (`Schwarzschild radius') and no ergosphere exists. For $a=M$ one
has $r_+=M$ and the ergosphere extends up to $r_0(\pi/2)=2M$ in the
equatorial plane. The values of the characteristic radii $r_{\rm ph}$,
$r_{\rm mb}$ and $r_{\rm ms}$ discussed above are given in Table 1.
\begin{table}
\caption{Photon orbit, marginally bound orbit and marginally stable orbit
for the Schwarzschild black hole and for the extreme Kerr black hole. In the
latter case one has to distinguish between direct and retrograde orbits.}
\medskip
\begin{tabular}{lrrr} \hline
                   & $r_{\rm ph}$ & $r_{\rm mb}$     & $r_{\rm ms}$ \\ \hline
$a=0$              & 3M           & 4M                    & 6M \\
$a=M$ (direct)     & M            & M                     & M  \\
$a=M$ (retrograde) & 4M           & $(3+2\sqrt{2})$M      & 9M \\ \hline
\end{tabular}
\end{table}
In the extreme case $r_{\rm ph}^{(d)}$, $r_{\rm mb}^{(d)}$ and
$r_{\rm ms}^{(d)}$ [the $(d)$ indicates direct orbits] all coincide with
$r_+=M$. However, defining proper radial distances according to
\begin{equation}
\delta(r_1,r_2)=\int\limits_{r_1}^{r_2}\sqrt{g_{rr}}\,dr,\quad
g_{rr}=\frac{\Sigma}{\Delta}\quad ,
\label{prop}
\end{equation}
one finds \cite{bpt}
\begin{equation}
\lim_{a\to M} \delta(r_+,r_{\rm ph}^{(d)})=\frac{M}{2}\ln 3, \quad
\lim_{a\to M} \delta(r_+,r_{\rm mb}^{(d)})=M\ln(1+\sqrt{2})
\end{equation}
\begin{equation}
\mbox{and}\quad  \lim_{a\to M} \delta(r_+,r_{\rm ms}^{(d)})=\infty.
\end{equation}
Moreover, because of the double zero of $\Delta$ at $r=M$ for the extreme
Kerr metric, one obtains
\begin{equation}
\delta(M,r)=\infty \quad \mbox{for any} \quad r>M.
\end{equation}
This means, the horizon (as well as $r_{\rm ph}^{(d)}$, $r_{\rm mb}^{(d)}$ and
$r_{\rm ms}^{(d)}$ !) has an infinite proper radial distance from any point
in the `exterior' region $r>M$. On the other hand, the proper time of an
infalling particle starting at some $r>M$ needed
to reach the horizon remains finite.
\begin{figure}
\scalebox{0.8}{\includegraphics{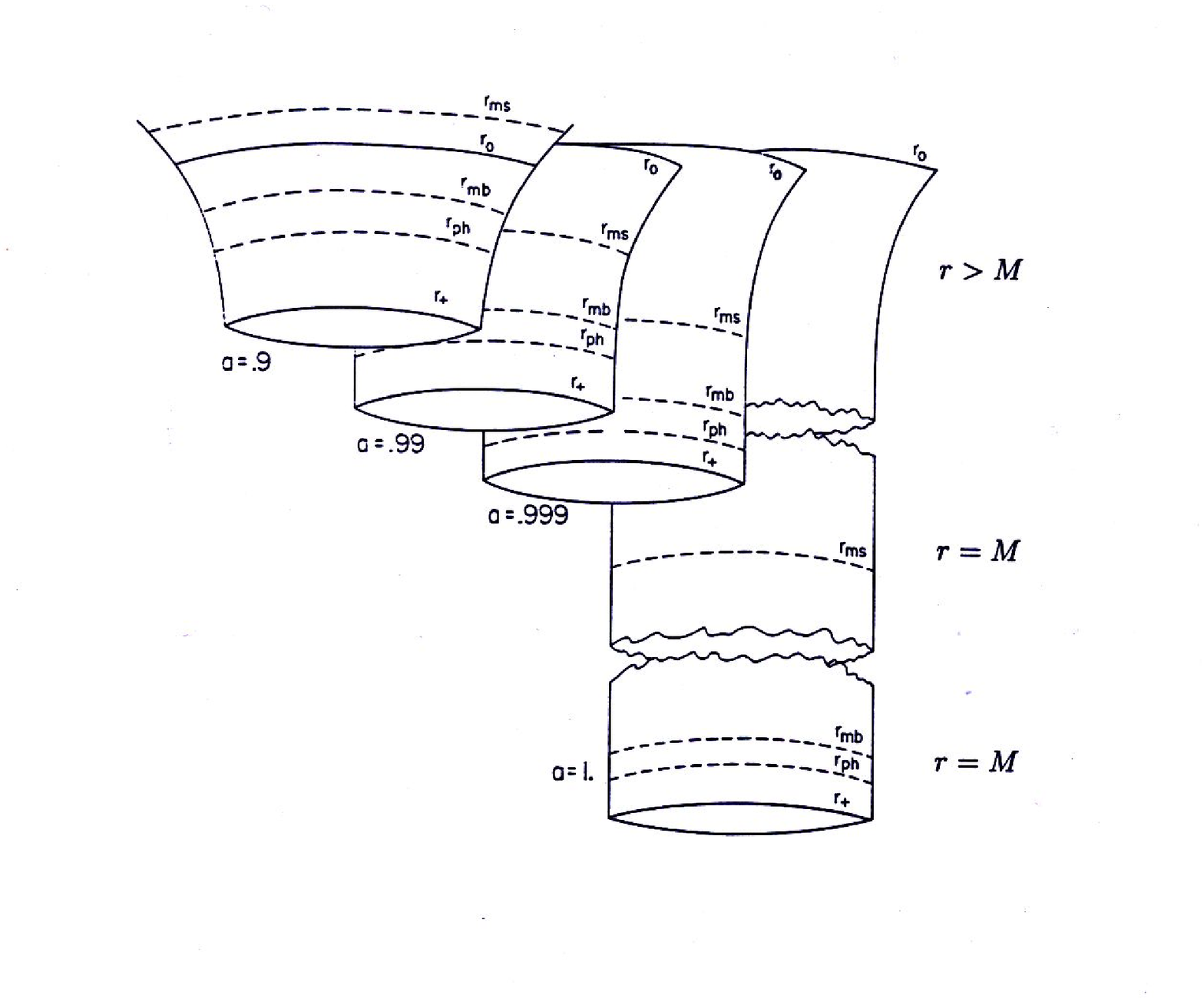}}
\caption{Embedding diagrams of the $r\ge r_+$ part of
the `plane' $\vartheta=\pi/2$,
$t={\rm constant}$ of the Kerr metric approaching the limit $a=M$ (from
Bardeen {\it et al.} \cite{bpt}; $a$ is given in units of $M$).
The positions of the direct orbits
$r_{\rm ph}$, $r_{\rm mb}$, $r_{\rm ms}$ and of the boundary $r_0$ of the
ergosphere  are shown.}
\label{fig1}
\end{figure}
The geometrical situation can nicely be illustrated by embedding the
$r\ge r_+$ part of the `plane'
$\vartheta=\pi/2$, $t={\rm constant}$ into three--dimensional Euclidean space
\cite{bpt}, see Fig.~\ref{fig1}.
In the limit $a=M$ an infinitely long `throat' characterized by $r=M$
(circumference: $4\pi M$) appears. The horizon is situated at the bottom and
the direct orbits corresponding to $r_{\rm ph}^{(d)}$, $r_{\rm mb}^{(d)}$ and
$r_{\rm ms}^{(d)}$ are located at different places along the throat.
However, the proper time of an infalling particle needed for passing through this
throat, is zero. Bardeen and Horowitz \cite{bh} have studied the `throat
geometry' ($r=M$) by means of the coordinate transformation
\begin{equation}
\tilde{r}=\frac{r-M}{\lambda}, \quad \tilde{\vartheta}=\vartheta, \quad
\tilde{\varphi}=\varphi-\Omega_{\rm H}t, \quad \tilde{t}=\lambda t
\end{equation}
in the limit $\lambda\to 0$. Note that
\begin{equation}
\omega(r_+,\vartheta)=\Omega_{\rm H}={\rm constant}
\label{oh}
\end{equation}
defines the `angular velocity of the horizon'. For the extreme Kerr black
hole it is given by
\begin{equation}
\Omega_{\rm H}=\frac{1}{2M}.
\end{equation}
One obtains
\begin{equation}
ds^2=M^2(1+\cos^2\tilde{\vartheta})\left(\frac{d\tilde{r}^2}
{\tilde{r}^2}+d\tilde{\vartheta}^2-\frac{\tilde{r}^2d\tilde{t}^2}{4M^4}\right)
+\frac{4M^2\sin^2\tilde{\vartheta}}{1+\cos^2\tilde{\vartheta}}
\left(d\tilde{\varphi}+\frac{\tilde{r}d\tilde{t}}{2M^2}\right)^2.
\label{kt}
\end{equation}
This represents a completely nonsingular vacuum solution of Einstein's
equations\footnote{The solution coincides with the metric given by
equations (56--59) of \cite{m98} and it belongs to a class of solutions
presented by Ernst \cite{e77}.}
which is geodesically complete but no longer asymptotically flat \cite{bh}.
The area
of all surfaces $\tilde{r}={\rm constant}$, $\tilde{t}={\rm constant}$ is
$8\pi M^2$ and equal to the area of the horizon of the extreme Kerr
black hole. In addition to $\partial/\partial\tilde{t}$ and
$\partial/\partial\tilde{\varphi}$ the metric (\ref{kt})
has two more Killing fields
\cite{w, bh}:
\begin{equation}
\left(\frac{\tilde{t}^2}{2}+\frac{1}{8\tilde{r}^2\Omega_{\rm H}^4}\right)
\frac{\partial}{\partial\tilde{t}}-\tilde{r}\tilde{t}\frac{\partial}
{\partial\tilde{r}}-\frac{1}{2\tilde{r}\Omega_{\rm H}^2}\frac{\partial}
{\partial\tilde{\varphi}},\quad \tilde{t}\frac{\partial}{\partial\tilde{t}}
-\tilde{r}\frac{\partial}{\partial\tilde{r}}.
\end{equation}
\section{Idealized routes to black holes}
A famous idealized route to the Schwarzschild black hole is the spherically
symmetric dust collapse (`Oppenheimer--Snyder collapse') \cite{t, os}.
Bardeen and Wagoner \cite{bw} have shown approximately that a uniformly
rotating disk of dust allows a quasi--stationary transition (parametric
collapse) towards the extreme Kerr black hole. This has been confirmed
by the exact solution to the disk problem \cite{nm1, nm2}, see also \cite{m98}.
Here the main results will be reviewed.

The line element for any stationary, axisymmetric, purely rotating
perfect--fluid configuration may be written in Lewis--Papapetrou coordinates
as
\begin{equation}
ds^2=e^{2\alpha}(d\varrho^2+d\zeta^2) + W^2e^{-2\nu}
(d\varphi-\omega\, dt)^2-e^{2\nu}dt^2.
\label{lin1}
\end{equation}
The four functions $\alpha$, $W$, $\nu$ and $\omega$ depend on $\varrho$ and
$\zeta$ only. Along the symmetry axis, $\varrho=0$, the condition of elementary
flatness requires
\begin{equation}
\lim\limits_{\varrho\to 0}\frac{W}{\varrho}e^{-(\alpha+\nu)} = 1
\end{equation}
and at infinity ($\varrho^2+\zeta^2\to \infty$) the asymptotic line element
has to take the Minkowski form in cylindrical coordinates,
\begin{equation}
ds^2=d\varrho^2+d\zeta^2+\varrho^2d\varphi^2-dt^2,
\end{equation}
\begin{equation}
\mbox{i.e.,}\quad
\alpha\to 0, \quad W\to\varrho, \quad \nu\to 0, \quad \omega\to 0.
\end{equation}
An equivalent form of (\ref{lin1}) is
\begin{equation}
ds^2=e^{-2U}\left[e^{2k}(d\varrho^2+d\zeta^2)+W^2 d\varphi^2\right]
-e^{2U}(dt+A\, d\varphi)^2
\label{lin2}
\end{equation}
\begin{equation}
\mbox{with}\quad \alpha=k-U, \quad W^{-1}e^{2\nu}\pm\omega=
\left(W\, e^{-2U}\mp A\right)^{-1}.
\end{equation}
For the disk of dust one obtains
\begin{equation}
W\equiv \varrho
\end{equation}
and the whole problem can be formulated as a boundary value problem \cite{nm1}
of the Ernst equation
\begin{equation}
(\Re f)(f,_{\rho \rho}+f,_{\zeta\zeta} + \rho^{-1}f,_{\rho})
=f,_{\rho}^2+f,_{\zeta}^2.
\label{ernst}
\end{equation}
The metric can easily be obtained from the complex Ernst potential
\begin{equation}
f=e^{2U}+ib
\end{equation}
whose real part is equal to $e^{2U}$. The functions $A$ and $k$ can be
calculated by integration:
\begin{equation}
A(\varrho,\zeta)=\int\limits_0^\rho \tilde{\rho}
e^{-4U}b,_{\zeta} d\tilde{\rho},
\quad
k(\varrho,\zeta)=
\int\limits_0^\rho \tilde{\rho}\left[U,_{\tilde{\rho}}^2-U,_{\zeta}^2+
\frac{e^{-4U}}{4}(b,_{\tilde{\rho}}^2-b,_{\zeta}^2)\right] d\tilde{\rho}.
\end{equation}
[In the integrands, one has $U=U(\tilde{\rho},\zeta)$ and $b=b(\tilde{\rho},
\zeta)$.] Like the Kerr metric, the solution for a uniformly rotating disk
of dust depends on two parameters, the total mass $M$ and the angular
momentum $J$. In contrast to (\ref{jmk}) the relation
\begin{equation}
J\ge M^2
\end{equation}
holds for the disk of dust. Instead of $M$ and $J$ other parameter pairs can
be used. In the following, the solution is represented in terms of the two
parameters $\varrho_0$ (coordinate radius of the disk) and $\mu$, where
$\mu$ is related
to $\Omega$ (angular velocity of the disk) by
\begin{equation}
\mu=2\Omega^2\varrho_0^2e^{-2V_0};\quad V_0\equiv U(\varrho=0,\zeta=0).
\label{mu}
\end{equation}
The Ernst potential is given by hyperelliptic integrals \cite{nm2}:
\begin{equation}
f=\exp\left\{\int\limits_{K_1}^{K_a}\frac{K^2dK}{Z} +
\int\limits_{K_2}^{K_b}\frac{K^2dK}{Z} - v_2\right\},
\label{f}
\end{equation}
with
\begin{equation}
Z=\sqrt{(K+iz)(K-i\bar{z})(K^2-K_1^2)(K^2-K_2^2)}, \quad z=\varrho+i\zeta,
\end{equation}
\begin{equation}
K_1=-\bar{K}_2=\rho_0\sqrt{\frac{i-\mu}{\mu}} \quad (\Re K_1<0,\,\,
\mbox{a bar denotes complex conjugation}).
\label{K12}
\end{equation}
The upper integration limits $K_a$ and $K_b$ in (\ref{f})
have to be calculated from
\begin{equation}
\int\limits_{K_1}^{K_a}\frac{dK}{Z} +
\int\limits_{K_2}^{K_b}\frac{dK}{Z} = v_0, \quad
\int\limits_{K_1}^{K_a}\frac{KdK}{Z} +
\int\limits_{K_2}^{K_b}\frac{KdK}{Z} = v_1,
\label{jacobi}
\end{equation}
where the functions $v_0$, $v_1$ and $v_2$
in (\ref{jacobi}) and (\ref{f}) are
given by
\begin{equation}
v_0=\int\limits_{-i\rho_0}^{+i\rho_0}\frac{H}{Z_1}dK,\quad
v_1=\int\limits_{-i\rho_0}^{+i\rho_0}\frac{H}{Z_1}KdK,\quad
v_2=\int\limits_{-i\rho_0}^{+i\rho_0}\frac{H}{Z_1}K^2dK,
\label{v}
\end{equation}
\begin{equation}
H=\frac{\mu\ln\left[\sqrt{1+\mu^2(1+K^2/\rho_0^2)^2} + \mu(1+K^2/\rho_0^2)
\right]}{\pi i \rho_0^2\sqrt{1+\mu^2(1+K^2/\rho_0^2)^2}} \qquad (\Re H =0),
\end{equation}
\begin{equation}
Z_1=\sqrt{(K+iz)(K-i\bar{z})} \qquad (\Re Z_1<0 \,\, \mbox{for $\rho$, $\zeta$
outside the disk}).
\label{r}
\end{equation}
In (\ref{v}) one has to integrate along the imaginary axis. The
integrations from $K_1$ to $K_a$ and from $K_2$ to $K_b$ in (\ref{f}) and
(\ref{jacobi}) have to
be performed along the same paths in the two--sheeted Riemann surface associated
with $Z(K)$. The problem of finding $K_a$ and $K_b$ from
(\ref{jacobi}) is a special case of Jacobi's inversion problem.
It generalizes the
definition of elliptic functions and can be solved in terms of hyperelliptic
theta functions (\cite{goe, ro},  see also \cite{stahl, kra, bob}).
Using a formula for Abelian integrals of the third kind derived by Riemann
(see \cite{stahl}) it is also possible to express the Ernst potential $f$
directly in terms of theta functions \cite{nkm}. On the symmetry axis ($\rho=0$)
and in the plane
of the disk ($\zeta=0$) all integrals in (\ref{f}) and (\ref{jacobi})
reduce to elliptic ones \cite{nm3}.

The solution has a positive surface mass--density \cite{nm3} and it is regular
everywhere outside the disk -- provided
\begin{equation}
0<\mu<\mu_0=4.62966184\dots
\end{equation}
(for $\mu>\mu_0$ one or more singular rings appear in the plane $\zeta=0$,
outside the disk). The parameter $V_0$ [see (\ref{mu})] depends on $\mu$
alone \cite{nm3}:
\begin{equation}
V_0=-\frac{1}{2}\sinh ^{-1}\left\{\mu+\frac{1+\mu^2}
{\wp[I(\mu);\frac{4}{3}\mu^2-4,\frac{8}{3}\mu(1+\mu^2/9)]-\frac{2}{3}\mu}
\right\},
\label{V0}
\end{equation}
\begin{equation}
I(\mu)=\frac{1}{\pi}\int\limits_0^{\mu}\frac{\ln(x+\sqrt{1+x^2})dx}
{\sqrt{(1+x^2)(\mu-x)}}
\end{equation}
with the Weierstrass function $\wp$ defined by
\begin{equation}
\int\limits_{\wp(x;g_2,g_3)}^{\infty}
\frac{dt}{\sqrt{4t^3-g_2t-g_3}}=x.
\end{equation}
The range $0<\mu<\mu_0$ corresponds to
$0>V_0>-\infty$. [$\mu_0$ is the first zero of the
denominator in (\ref{V0}).]
From (\ref{mu}) and (\ref{V0}) one
obtains the relation $\Omega(\mu,\rho_0)$.
\begin{figure}
\hspace{2cm}\scalebox{1.0}{\includegraphics{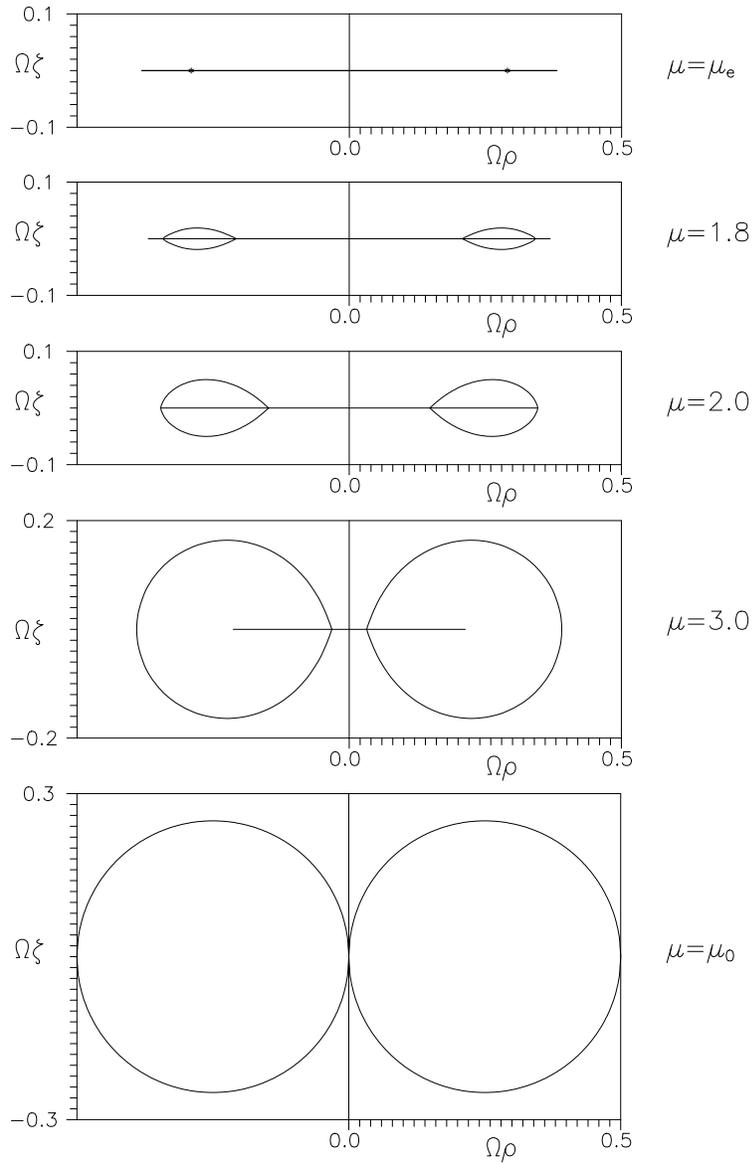}}
\caption{Formation of the ergosphere (from \cite{mk}). It appears for
$\mu>\mu_{\rm e}=1.68849\dots$ and has a toroidal shape. The horizontal
line represents the disk. For $\mu\to\mu_0$ the ergosphere is that
of the extreme Kerr black hole.}
\label{fig2}
\end{figure}

Whereas $\mu\ll 1$ characterizes the Newtonian limit (the `Maclaurin disk'),
$\mu\to\mu_0$ leads to the announced black hole limit:

\noindent For $\mu\to\mu_0$ one obtains
\begin{equation}
\frac{M^2}{J}\to 1, \quad 2\Omega M\to 1, \quad \Omega\varrho_0\to 0,\quad
\mbox{which means, for finite $M$,}
\end{equation}
\begin{equation}
\varrho_0\to 0.
\end{equation}
For the further discussion, we introduce spherical--like coordinates $R$,
$\vartheta$:
\begin{equation}
\varrho=R\,\sin\vartheta, \quad \zeta=R\,\cos\vartheta.
\label{R}
\end{equation}
The disk ($\zeta=0$, $\varrho\le\varrho_0$) shrinks to the origin $R=0$ of the
coordinate system.
For $R\ne 0$, the disk metric becomes exactly the $r>M$ part of
the extreme Kerr metric given by (\ref{k}) with $a=M$ ($\Omega_{\rm H}=
\Omega=1/2M$) and
\begin{equation}
r=R+M.
\label{rR}
\end{equation}
This can be shown as follows \cite{m98, m00}:

Let us first rewrite (\ref{f}) and
(\ref{jacobi}) in the equivalent form
\begin{equation}
f=\exp\left\{\,\int\limits_{K_b}^{K_a}\frac{K^2dK}{Z} - \tilde{v}_2\right\},
\quad \int\limits_{K_b}^{K_a}\frac{dK}{Z}=\tilde{v}_0, \quad
\int\limits_{K_b}^{K_a}\frac{KdK}{Z}=\tilde{v}_1,
\end{equation}
with
\begin{equation}
\tilde{v}_n=v_n - \int\limits_{K_1}^{K_2}\frac{K^ndK}{Z} \quad (n=0,1,2).
\end{equation}
($K_b$ is now on the other sheet of the Riemann surface.) In the limit
$\mu\to\mu_0$ one obtains for $R>0$, using (\ref{mu}) and (\ref{V0}),
\begin{equation}
\tilde{v}_0=\frac{2\Omega}{R} - \frac{\pi i\cos\vartheta}{2R^2},
\quad \tilde{v}_1=-\frac{\pi i}{2R}, \quad \tilde{v}_2=0
\end{equation}
(modulo periods).
In the above integrals from $K_b$ to $K_a$,
$Z$ can be replaced by $Z=K^2\sqrt{(K+iz)(K-i\bar{z})}$
since $K_1$ and $K_2$ both tend to zero [cf.~(\ref{K12})]. Hence, all integrals
become elementary and the unique result is
\begin{equation}
f=\frac{2\Omega R-1-i\cos \vartheta}{2\Omega R +1-i\cos \vartheta} \qquad
(R > 0),
\label{ek}
\end{equation}
which is the Ernst potential of the extreme Kerr solution. Note that $R=0$
($r=M$) characterizes the horizon (and the throat) of the extreme Kerr
black hole.

In Fig.~\ref{fig2} the formation of the ergosphere is shown \cite{mk}.
In the limit $\mu\to\mu_0$ the
ergosphere of the extreme Kerr black hole appears. Indeed, from (\ref{ergo}),
(\ref{R}) and (\ref{rR}) one finds the following equation
for the boundary of the extreme Kerr ergosphere in the (Weyl)
coordinates $\varrho$, $\zeta$ (note that $2M=1/\Omega$):
\begin{equation}
\left(\frac{\varrho_{\rm e}}{2M}-\frac{1}{4}\right)^2
+\left(\frac{\zeta_{\rm e}}{2M}\right)^2=\frac{1}{16}.
\end{equation}
\begin{figure}
\scalebox{0.7}{\includegraphics{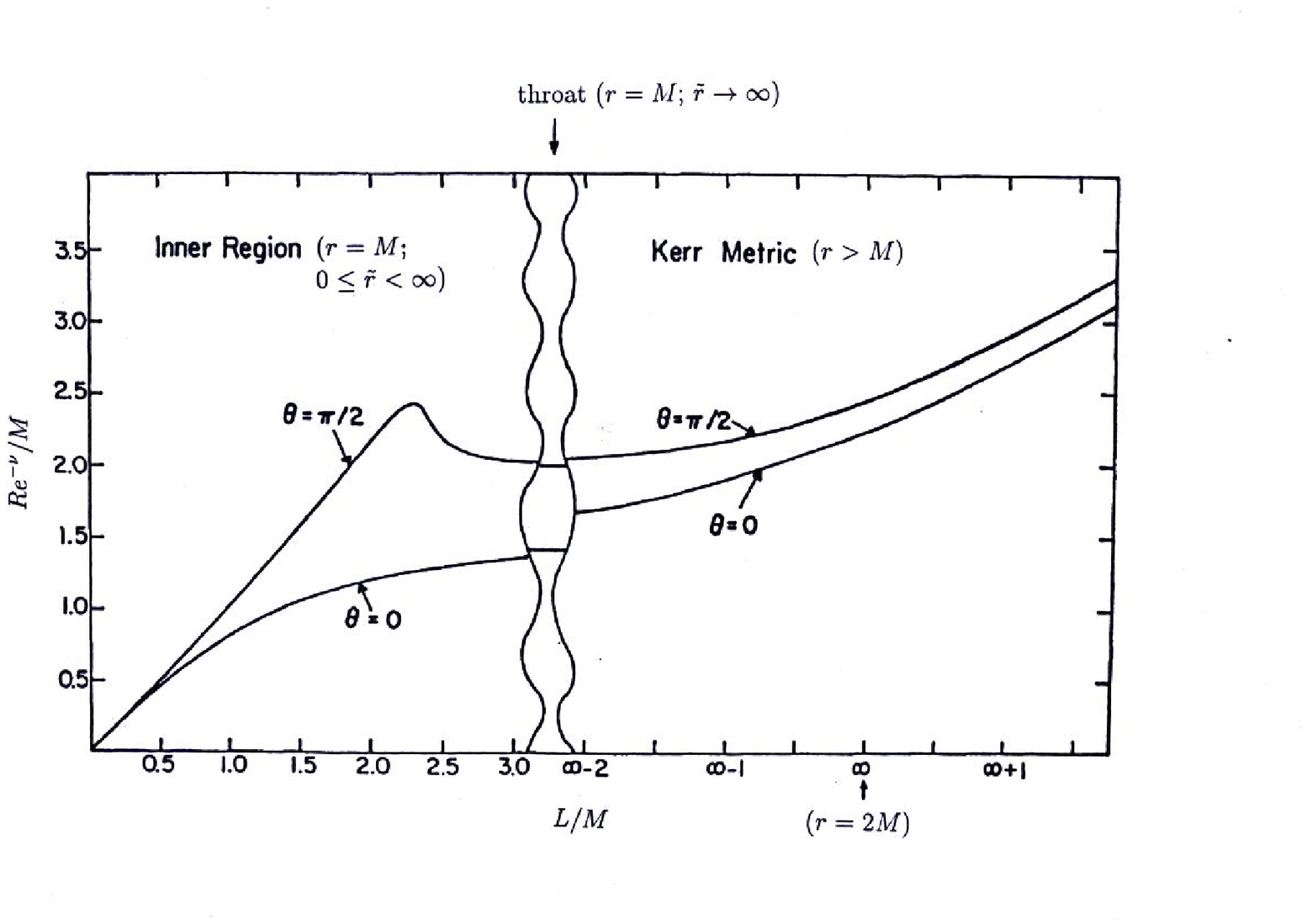}}
\caption{The $\mu\to\mu_0$ disk metric (from Bardeen and Wagoner \cite{bw},
slightly modified).
The ordinate shows the values of the function $R\,e^{-\nu}/M$
for $\vartheta=0$ (axis of symmetry) and for $\vartheta=\pi/2$
(equatorial plane). The abscissa shows $L/M$ where $L$ is
the proper radial distance from the center of
the disk. All points in the `exterior' ($r>M$) region have an infinite proper
distance to the `inner' ($r=M$) region which contains the  disk and its
surroundings.
The point `$L/M=\infty$' corresponds to the coordinate value $r=2M$,
`$L/M=\infty+x$'
means $\delta(2M,r)=xM$, cf.~Eq.~(\ref{prop}).}
\label{fig3}
\end{figure}
A completely different limit of the space--time, for $\mu\to\mu_0$,
is obtained for finite values of $R/\rho_0$ (corresponding
to the previously excluded $R=0$). Therefore, we consider a coordinate
transformation \cite{bw}
\begin{equation}
\tilde{r}=e^{-V_0}R,\quad \tilde{\vartheta}=\vartheta,
\quad \tilde{\varphi}=\varphi-\Omega t, \quad \tilde{t}=e^{V_0}t.
\label{co}
\end{equation}
(Note that finite $R/\rho_0$ correspond to finite $\tilde{r}$ in the limit.)
For $\mu<\mu_0$, this is nothing but the transformation to the corotating
system combined with a rescaling of $R$ and $t$. The
transformed Ernst potential $\tilde{f}$ is related to the Ernst
potential $f'$ in the corotating system
($R'=R$, $\vartheta'=\vartheta$,
$\varphi'=\varphi-\Omega t$, $t'=t$) according to
$\tilde{f}=f'\exp(-2V_0)$, i.e.,
\begin{equation}
\frac{\tilde{f}}{\tilde{r}^2}=\frac{f'}{R^2} \quad \mbox{for} \quad \mu<\mu_0.
\label{ff}
\end{equation}
However, for $\mu\to\mu_0$, the solutions $f'$ (finite $R>0$)
and $\tilde{f}$ (finite $\tilde{r}$) separate from each other.
(A similar phenomenon has been observed by Breitenlohner {\it et al.} for
some limiting solutions of the static, spherically symmetric
Einstein--Yang--Mills--Higgs equations \cite{bfm}.)
For finite $R>0$,
the extreme Kerr solution arises, while finite $\tilde{r}$ lead to
a solution which still describes a disk, with finite coordinate radius
\begin{equation}
\tilde{\varrho}_0=\lim\limits_{\mu\to\mu_0}(e^{-V_0}\varrho_0)=
\sqrt{2\mu_0}\,M.
\end{equation}
Note that the proper radius of the disk remains finite in the limit $\mu\to\mu_0$
as well. Its circumference is $4\pi M\sqrt{\mu_0/2-1}$ which is larger than
the circumference $4\pi M$ of the extreme Kerr throat.
The metric corresponding to $\tilde{f}$ (which can
be expressed in terms of theta functions) is regular
everywhere outside the disk, but it is not asymptotically flat.
The space--time structure of
both solutions ($f'$ and $\tilde{f}$) coincides at $R\to 0$ (the throat)
and $\tilde{r}\to\infty$ (spatial infinity). The relation (\ref{ff}) survives
in the form
\begin{equation}
\lim\limits_{\tilde{r}\to\infty}\,\frac{\tilde{f}}{\tilde{r}^2}=
\lim\limits_{R\to 0}\,\frac{f'}{R^2} \quad \mbox{as} \quad \mu\to\mu_0.
\end{equation}
[The limits have to be taken consistently with (\ref{co}).]
The Ernst potential $f'$ of the extreme Kerr solution
in the corotating system reads
\begin{equation}
f'=-\Omega^2R^2\left[\frac{2(1+i\cos\vartheta)^2}{2\Omega R +1-i\cos \vartheta}
+\sin^2\vartheta\right].
\end{equation}
Accordingly, for $\mu=\mu_0$ and $\tilde{r}\to\infty$,
\begin{equation}
\tilde{f}\to\tilde{f}_{\rm as} = -\Omega^2\tilde{r}^2\left[
\frac{2(1+i\cos\tilde{\vartheta})^2}{1-i\cos \tilde{\vartheta}}
+\sin^2\tilde{\vartheta}\right].
\end{equation}
Note that $\tilde{f}_{\rm as}$ belongs to the family of solutions to
the Ernst equation of the type $f=R^kY_k(\cos \vartheta)$
presented by Ernst \cite{e77}. The corresponding
{\it asymptotic} line element is given by the extreme Kerr throat
geometry (\ref{kt}). These exact results \cite{m98} confirm the
picture of the extreme relativistic limit of the rotating disk
developed by Bardeen and Wagoner \cite{bw}. Fig.~\ref{fig3} (taken from
\cite{bw}) combines
the two limits of the disk space--time for $\mu\to\mu_0$. The abscissa
characterizes the proper radial distance $L$ from the center of the disk:
\begin{equation}
L=\int\limits_0^{\tilde{r}}\sqrt{g_{\tilde{r}\tilde{r}}}\, d\tilde{r}=
\int\limits_0^R\sqrt{g_{RR}}\, dR.
\end{equation}
The two `worlds' are separated by the `throat region' which plays the role of
infinity for the `inner region'. From the point of view of the `exterior region'
(the $r>M$ part of the extreme Kerr metric)
it is the extreme Kerr throat as discussed in Section 2. Note
that the plotted quantity $R\,e^{-\nu}/M$, for $\vartheta=\pi/2$, is equal to
$\sqrt{g_{\varphi\varphi}}/M$, which is just the circumferential
radius\footnote{The circumferential radius
is defined as the circumference
divided by $2\pi$.} of a circle in the equatorial plane (concentrical to the
disk) divided
by $M$. The value of 2 in the throat region corresponds to the throat's
circumference $4\pi M$.
\begin{figure}
\scalebox{1}{\includegraphics{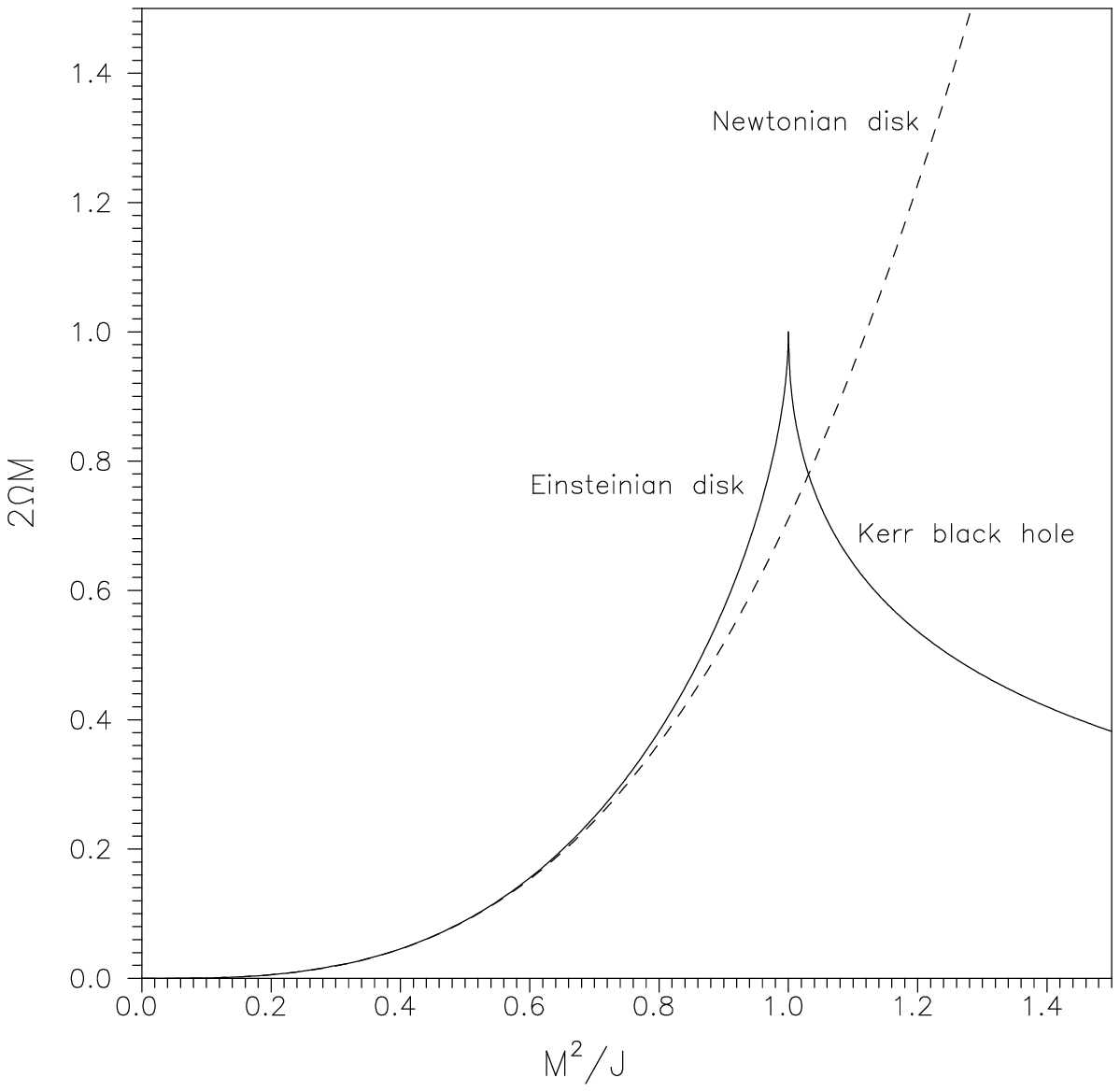}}
\caption{Parametric collapse of the rigidly rotating disk of dust (from
\cite{nm1}). The diagram shows the relation between $2\Omega M$ and
$M^2/J$ for the disk of dust and for the Kerr black hole
[$\Omega=\Omega_{\rm H}$, cf.~Eq.~(\ref{oh})]. The dashed line
corresponds to the
Newtonian disk solution (the Maclaurin disk) where $2\Omega M=
(9\pi^2/125)(M^2/J)^3$.}
\label{fig4}
\end{figure}

The quasi--stationary transition from the disk to the
Kerr black hole is illustrated in Fig.~\ref{fig4} where $2\Omega M$ is shown in
dependence on the parameter $M^2/J$ \cite{nm1}. The limit $M^2/J\to 1$ (from
the left) corresponds to the limit $\mu\to\mu_0$ discussed above. The `exterior'
metric becomes the $r>r_+$ part of the extreme Kerr metric. For $M^2/J>1$ no
disk solution exists and a genuine Kerr black hole forms. This transition
(parametric collapse) from the disk to the Kerr black hole is
{\it continuous} in the `exterior region'. In particular, all multipole moments
behave continuously, see \cite{kmn}.
\section{From stars to black holes}
The exterior metric of a spherically symmetric star, even in the case of
(dynamic) collapse, is always the Schwarzschild metric. This is a consequence
of Birkhoff's theorem \cite{birk}. Therefore, the spherically symmetric
collapse of a sufficiently massive, non--rotating star at the end of its
life leads quite naturally to a Schwarzschild black hole, as in the
idealized Oppenheimer--Snyder case. On the other hand, a continuous
quasi--static transition from stars to black holes is not possible. The
surface of a star cannot be identical with the horizon of a black hole
since the horizon is a null--hypersurface. Under the quite reasonable
assumption that the mass--energy density does not increase outwards in the
star, one can show that the radius $r_*$ of a spherically symmetric, static
star (in Schwarzschild coordinates) always satisfies
\begin{equation}
r_*>\frac{9}{8}\,r_+,
\end{equation}
where $r_+=2M$ is the corresponding Schwarzschild radius (see \cite{wei},
for example). Thus, at most the $r>(9/8)r_+$ part of the Schwarzschild
vacuum metric is relevant outside static stars and the black hole state can
only be reached dynamically.

For rotating stars, the situation is different in both previously mentioned
respects. Firstly, the exterior metric is not the Kerr metric in general.
(There is no analogue to Birkhoff's theorem.) It is general belief, based on
Penrose's cosmic censureship conjecture combined with the black hole
uniqueness theorems, cf.~\cite{he}, that the (dynamic) collapse of a rotating
star leads asymptotically ($t\to\infty$) to the Kerr black hole, i.e., to the
Kerr metric outside the horizon. This has not yet been proved, however.
But secondly, a continuous quasi--stationary route from rotating stars
to rotating black holes via the extreme Kerr metric seems possible --- as in
the idealized (and certainly unstable) case of the rigidly rotating disk
of dust. The problem of the impossible identity of the star's surface with
the horizon would be circumvented by the `throat region', and the whole
$r>r_+$ part of the extreme Kerr metric could be relevant in the
`exterior region', as discussed above. Results on differentially
rotating disks of dust \cite{am, a} give some evidence that the extreme Kerr
limit is a generic possibility. General relativistic, rotating stellar
models can only be treated by numerical methods so far. However, the recently
achieved progress in accuracy by many orders of magnitude \cite{akm} justifies
the hope of finding a more realistic, quasi--stationary route to the Kerr
metric.

\vspace*{0.25cm} \baselineskip=10pt{\small \noindent I would like to thank
M.~Ansorg, A.~Kleinw\"achter, G.~Neugebauer and D.~Petroff for many valuable
discussions.}

\end{document}